\documentclass[a4paper]{spie}
\usepackage{graphicx}
\usepackage{latexsym}

\title{The Herschel/PACS 2560 bolometers imaging camera}

\author{Nicolas Billot\supit{a}, Patrick Agn\`{e}se\supit{b},
Jean-Louis Augu\`{e}res\supit{a}, Alain B\'{e}guin\supit{b}, Andr\'{e}
Bou\`{e}re\supit{a}, Olivier Boulade\supit{a}, Christophe
Cara\supit{a}, Christelle Clou\'{e}\supit{a}, Eric Doumayrou\supit{a},
Lionel Duband\supit{c}, Beno\^{i}t Horeau\supit{a}, Isabelle Le
Mer\supit{a}, Jean Le Pennec\supit{a}, J\'{e}rome Martignac\supit{a},
Koryo Okumura\supit{a}, Vincent Rev\'{e}ret\supit{d}, Marc
Sauvage\supit{a}, Fran\c{c}ois Simoens\supit{b} and Laurent
Vigroux\supit{e} \skiplinehalf \supit{a}Service d'Astrophysique,
DAPNIA, CEA Saclay, 91191 Gif sur Yvette, FRANCE; \\
\supit{b}Laboratoire Infra-Rouge, LETI, CEA Grenoble, 38054 Grenoble,
FRANCE; \\ \supit{c}Service des Basses Temp\'{e}ratures, DRFMC, CEA
Grenoble, 38054 Grenoble, FRANCE; \\ \supit{d}European Southern
Observatory, Vitacura, Casilla 19001 Santiago 19, CHILE; \\
\supit{e}Institut d'Astrophysique de Paris, 75014 Paris, FRANCE }

\authorinfo{Further author information: Nicolas Billot, e-mail:
nbillot@cea.fr\\PACS Photometer is funded by CNES and CEA.}

\begin{document} 
 \maketitle 

\begin{abstract}
The development program of the flight model imaging camera for the
PACS instrument on-board the Herschel spacecraft is nearing
completion. This camera has two channels covering the 60 to 210
microns wavelength range. The focal plane of the short wavelength
channel is made of a mosaic of 2x4 3-sides buttable bolometer arrays
(16x16 pixels each) for a total of 2048 pixels, while the long
wavelength channel has a mosaic of 2 of the same bolometer arrays for
a total of 512 pixels. The 10 arrays have been fabricated,
individually tested and integrated in the photometer. They represent
the first filled arrays of fully collectively built bolometers with a
cold multiplexed readout, allowing for a properly sampled coverage of
the full instrument field of view.  The camera has been fully
characterized and the ground calibration campaign will take place
after its delivery to the PACS consortium in mid 2006. The bolometers,
working at a temperature of 300~mK, have a NEP close to the BLIP limit
and an optical bandwidth of 4 to 5 Hz that will permit the mapping of
large sky areas. This paper briefly presents the concept and
technology of the detectors as well as the cryocooler and the warm
electronics. Then we focus on the performances of the integrated
focal planes (responsivity, NEP, low frequency noise, bandwidth).
\end{abstract}


\section{INTRODUCTION}
\label{sect:intro}

The Herschel Space Observatory is the third ``corner stone'' mission of
the European Space Agency. It will be launched by an Ariane 5 rocket
in the course of 2008. Herschel will be equipped with the largest
telescope ever sent in space (\O\,3.5m) and will carry out
spectroscopic and imaging observations in the 60 $\mu$m to 670 $\mu$m
wavelength range. Herschel's payload consists of three
instruments. (1) HIFI is a very high resolution heterodyne
spectrometer ($R\sim10^7$), (2) SPIRE is an imager and an imaging
spectrometer, operating in the 210-670 $\mu$m band, using spider-web
bolometers coupled to Winston cones (see ref.~\citenum{turner} for
details), and (3) PACS covers the 60-210 $\mu$m range and is both an
imaging spectrometer using photo-conducting detectors, and an imager
using novel technology bolometers described in this paper.

The main science objectives of Herschel are twofold. First, Herschel
will perform large scale surveys of nearby dark clouds, regions where
stars form, in order to identify the mechanisms responsible for the
distribution of stellar masses. Indeed we now realize that a star's
mass, when it enters the main sequence, is in fact determined when the
gas and dust cloud in which it will later form separates itself from
its parent cloud and starts to collapse. At this early stage, it is
mostly heated by the contraction and its temperature is such that it
radiates most of its energy in the Herschel band. By observing very
large numbers of these prestellar cores, we will shed light upon the
processes that lead to their formation and to their mass distribution.

Herschel will also peer into the distant Universe. Half of the
extragalactic background light reaches earth in the infrared, with a
peak in Herschel's bandpass. Herschel will perform deep surveys in
dark regions of the sky to identify and locate the galaxies
responsible for this background. This will allow the reconstruction
of the star formation history of the Universe during the last $\sim$10
Gyr. This star formation history is in fact the result of the galaxy
formation process, thus Herschel will participate in the construction
of a plausible scenario that leads from the very homogeneous Universe
of the Big Bang epoch to the highly structured Universe of galaxies
that we see now. For more detailed descriptions, see
ref.~\citenum{pilbratt} for the Herschel mission, and
ref.~\citenum{poglitsch} for the PACS instrument.

%%-----------------------------------------------------------
\section{Overview of PhFPU, the PACS photometer} 
\label{sect:overview}

The imaging part of PACS is referred to as the Photometer Focal Plane
Unit or PhFPU. It is designed for dual-band imaging in the range 60 to
210 $\mu$m. It consists of two channels: the ``Blue'' one covering the
range 60 to 130 $\mu$m, and the ``Red'' one from 130 to 210
$\mu$m. The split of wavelength is done with dichroic optics in
front of the photometer. The Blue channel is itself split into two
sub-ranges, from 60 to 85 $\mu$m and 85 to 130 $\mu$m by means of a
filter wheel. Observations are therefore performed simultaneously in
either one of the two blue bands plus the red band (see
figure~\ref{fig:phfpu} for a picture of the flight model of the
photometer).

\begin{figure}[htbp]
\begin{minipage}{0.49\linewidth}
\begin{center}
\begin{tabular}{c}
\includegraphics[width=6cm]{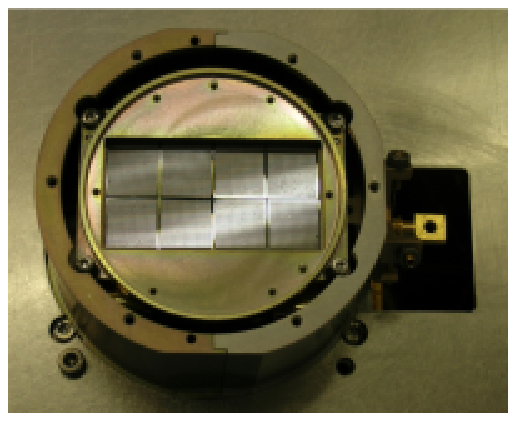}\end{tabular}
\end{center}
\end{minipage}
\hfill
\begin{minipage}{0.49\linewidth}
\begin{center}
\begin{tabular}{c}
\includegraphics[width=6cm]{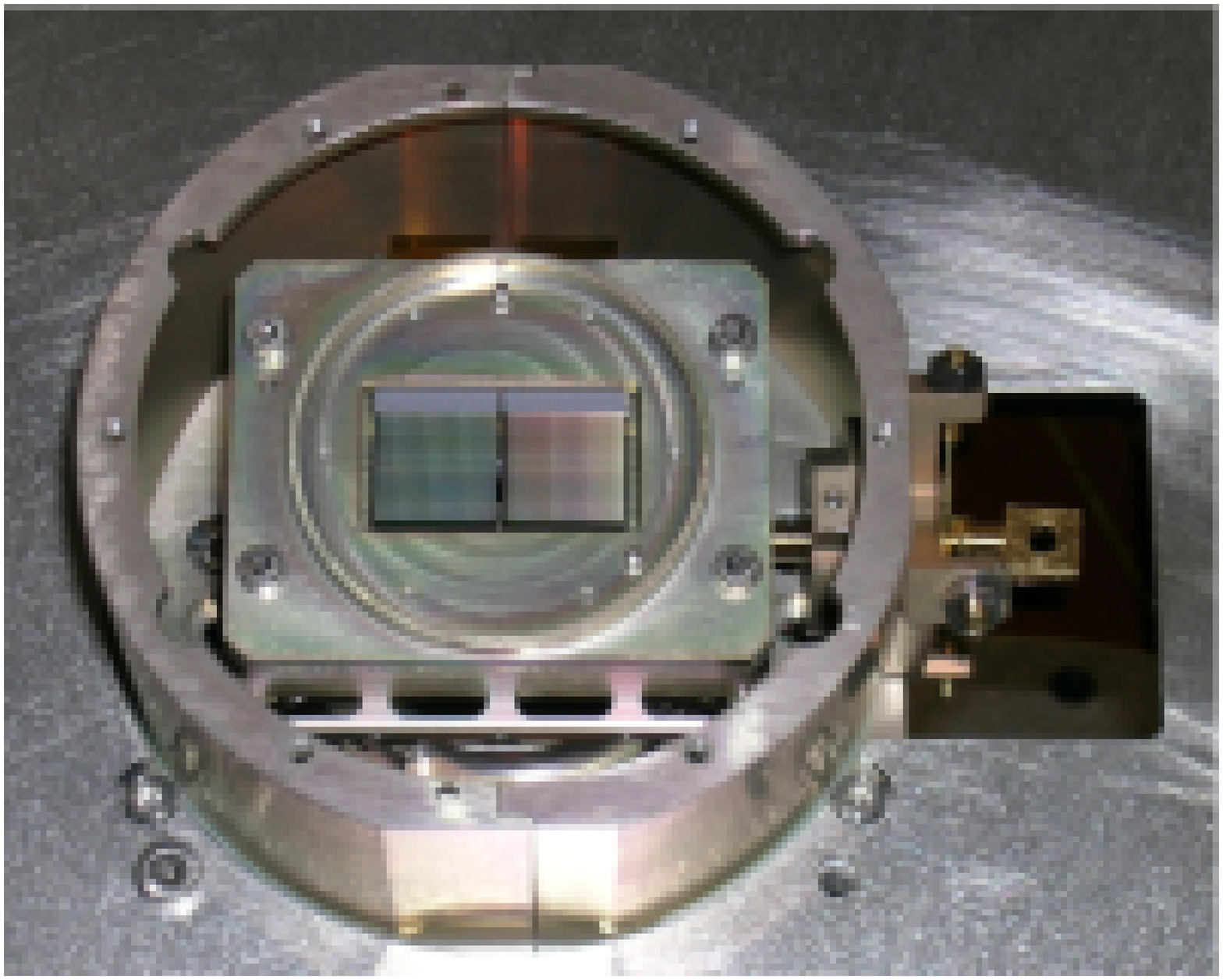}
\end{tabular}
\end{center}
\end{minipage}
\caption{\label{fig:bfps}The Blue (left: 64x32 pixels) and Red (right:
32x16 pixels) FM focal planes.}
\end{figure}

Each channel has a focal plane based on a mosaic of filled arrays of
bolometers (sect.~\ref{sect:bolometers}): 2048 pixels for the Blue
channel (arranged in a mosaic of 4x2 arrays of 16x16 pixels each), and
512 pixels for the Red channel (2 arrays of 16x16 pixels each). The
field of view is 3.5$'$x1.75$'$ for both channels and is fully sampled
by the filled arrays for the central
wavelengths. Figure~\ref{fig:bfps} shows the two focal planes of the
FM photometer. A cryocooler, based on an $^3$He sorption cooler, is
used to cool both focal planes to 300~mK
(sect.~\ref{sect:cryocooler}).

Each focal plane is mounted inside a structure connected to the 300~mK
stage. This structure is itself suspended inside the 2 K structure by
means of kevlar wires. A 300~mK filter is mounted on top of each
bolometer focal plane. The detectors and their cold readout
electronics at 300~mK are electrically connected at the 2 K stage to a
second level of electronics. Most of the power is dissipated at this
stage since the thermal budget at 300~mK is obviously very
tight. Table~\ref{tab:phfpu} summarizes the specifications of the
photometer.

\begin{figure}[htbp]
\begin{minipage}{0.62\linewidth}
\begin{center}       
\begin{tabular}{|c|c|c|}
\hline
Spectral range & \multicolumn{2}{c|}{60 -- 210 $\mu$m in two channels} \\
\hline
Field of view & \multicolumn{2}{c|}{3.5$'$x1.75$'$ per channel} \\
\hline
Image quality & \multicolumn{2}{c|}{Diffraction limited} \\
\hline
Operating temperature & \multicolumn{2}{c|}{$\sim$300 mK} \\
\hline
Thermal budget & \multicolumn{2}{c|}{ 10 $\mu$W at 300 mK} \\
\hline 
Autonomy & \multicolumn{2}{c|}{  46 h} \\
\hline
Channels & Blue & Red \\
 \hline
Central wavelength & 73 or 107 $\mu$m  & 166 $\mu$m \\
\hline
Bandwidth $\Delta \lambda$ & 33 or 43 $\mu$m  & 45 $\mu$m \\
\hline
Focal plane & 4x2 arrays & 2x1 arrays \\
\hline
Number of pixels & 2048 & 512 \\
\hline
Pixel field of view & 3.2 $''$ & 6.5 $''$ \\
\hline
\end{tabular}
\end{center}
\renewcommand\figurename{Table}
\caption{\label{tab:phfpu}Specifications for the PACS photometer.} 
\addtocounter{table}{1}
\addtocounter{figure}{-1}
\renewcommand\figurename{Figure}
\end{minipage}
\hfill
\begin{minipage}{0.36\linewidth}
\begin{center}
\begin{tabular}{c}
\includegraphics[width=5.7cm]{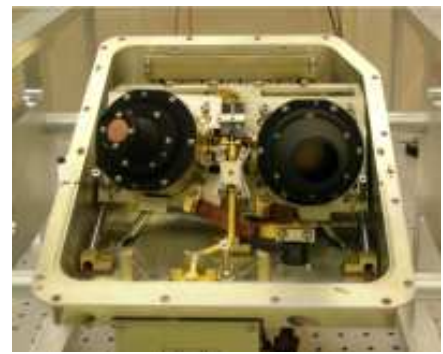}
\end{tabular}
\end{center}
\caption
{\label{fig:phfpu} The flight model of the PACS photometer being
assembled, Red channel is on the left and Blue on the right. PhFPU
dimensions are 260$\times$348.5$\times$216~mm and its weight is
8.2~kg.}
\end{minipage}
\end{figure} 

%%-----------------------------------------------------------
\section{The bolometer arrays} 
\label{sect:bolometers}

\subsection{Detection principle} 
\label{sect:principe}

The PACS bolometer arrays are based on the resonant absorption of the
sub-millimeter electromagnetic radiation. In this mode, an absorption
layer matched to vacuum impedance (377 $\Omega$/$\Box$) is located
above a reflector. In a classical point of view, standing waves
generated between incident and reflected radiations allow a
theoretical thermal absorption up to 100\% for a wavelength equal to
four times the distance between reflector and absorber. This principle
introduced by our group in 1996 for detection purposes, is now widely
applied in most recent bolometer developments.  The metal absorber is
deposited on a crystalline silicon mesh insulated from a heat sink by
four thin silicon rods (2$\mu$m x 5$\mu$m section - 600$\mu$m
long). The time constant of the detector is given by the heat capacity
of the insulated structure and the thermal conductance of the rods. To
lower as much as possible the heat capacity of the sensitive part of
the bolometer (the insulated structure), we act on both components of
the heat capacity: mass and specific heat.  Reduction of the mass is
obtained by thinning the silicon support to a limit compatible with
the process technology (5$\mu$m). Subsequent etching of the silicon
layer, to produce a mesh, achieves the mass reduction process.
Specific heat of crystalline materials drops rapidly when lowering
temperature below 1 K. We take advantage of this physical property by
running detectors below 300 mK. The specific heat of metals and
amorphous materials decreases less rapidly. A metal alloy in the
superconductor state (titanium nitride), far from the transition, is
chosen as absorber to bypass this drawback.  The temperature elevation
of the sensitive part must be measured. A semiconductor thermometric
structure is fitted out on the mesh. This structure is a thin
($<$1$\mu$m) and elongated silicon layer heavily doped with phosphorus
and 50\% compensated by boron ions, electrically insulated from the
mesh. This thermometer structure was measured to be the most
significant part of the heat capacity of the sensitive part of the
detector.

\subsection{Description of the bolometer arrays} 
\label{sect:arrays}
The way to collectively build large filled arrays while satisfying all
the prescriptions outlined above is to use two silicon chips: one
containing the absorbing insulated meshes with thermometers (the
pixels) and the other containing the reflectors and the cold readout
electronics. We then hybridize both with well defined indium bumps to
achieve the resonant cavity.  Double Silicon On Insulators (SOI)
wafers are used for detectors layers. Deep etching (400 $\mu$m) in one
direction, and surface etching (6$\mu$m) in the other produce meshes
and rods on the metallized silicon surface. Before that, surface
etching of the heavily doped silicon layer produces the mesa structure
of the thermometer. This structure is electrically insulated from the
mesh by the wafer SiO$_2$ upper insulation level. Detectors of large
sensitive surfaces are thus produced with sufficiently low heat
capacity to avoid any light concentrator as needed in classical
bolometric cameras. This solution opens the way for filled arrays
collective production.

The second chip to be hybridized is also a silicon integrated
circuit. It carries on its surface the gold reflectors covering C-MOS
readouts and multiplexing circuits. C-MOS (N\&P) transistors circuits
were adopted there to ensure electric functions at the detectors
operating temperature of 300 mK.  As the noise density for these
circuits is large, with respect to FET transistors, we decided to run
thermometers at very high resistance (in the T$\Omega$ range) to
ensure large signals. FET transistors, inefficient below 100 K, cannot
be used in our case. The disadvantage of T$\Omega$ circuits is the
difficulty to propagate signals on significant distances. The
proximity of the hybridized stages (a few millimeters) relaxes the
problem.

Chips including 16 x16 pixels and chips including 16 x16 MOS readout
circuits are manufactured for proper combination leading to sub-units
of 256 detectors. These sub-units are designed to be buttable on three
sides for large focal planes assembly.

The wavelength absorption requirement was initially ensured by two
sizes of indium bumps (20 and 25 $\mu$m), according to
calculations. Spectral reflection measurements with a Fourier
Transform Spectrometer showed that the shorter bump size is sufficient
to cope with both wavelength range requirements. Only the ``20 $\mu
m$'' type associated to the Blue channel was then manufactured.

A large focal plane containing 2560 bolometers is not really
compatible, in a space project, with a ``one readout channel per
detector'' policy and multiplexing is therefore mandatory. This
function is ensured by MOS transistors used here as gates at the
readout chip level. A 16 to 1 multiplexing is now currently achieved
reducing the total output channels to 160. Frames are then produced at
40 Hz.

The power dissipation available at the 300~mK level is very low
(10$\mu$W). The only way to meet this requirement is to output the
bolometer signals from the 300~mK readout stage into the M$\Omega$
range. When including the multiplexing to the frame frequency, the
available length range is of the order of ten centimeters. A second
readout stage (impedance adaptation) is then provided a few
centimeters away on a part of the focal plane linked to the satellite
2~K level. There, 3.5~mW of power is available and sufficient to
transfer signals to the warm electronics.

\begin{figure}[htbp]
\begin{minipage}{0.6\linewidth}
\begin{center}
\begin{tabular}{c}
\includegraphics[width=9cm]{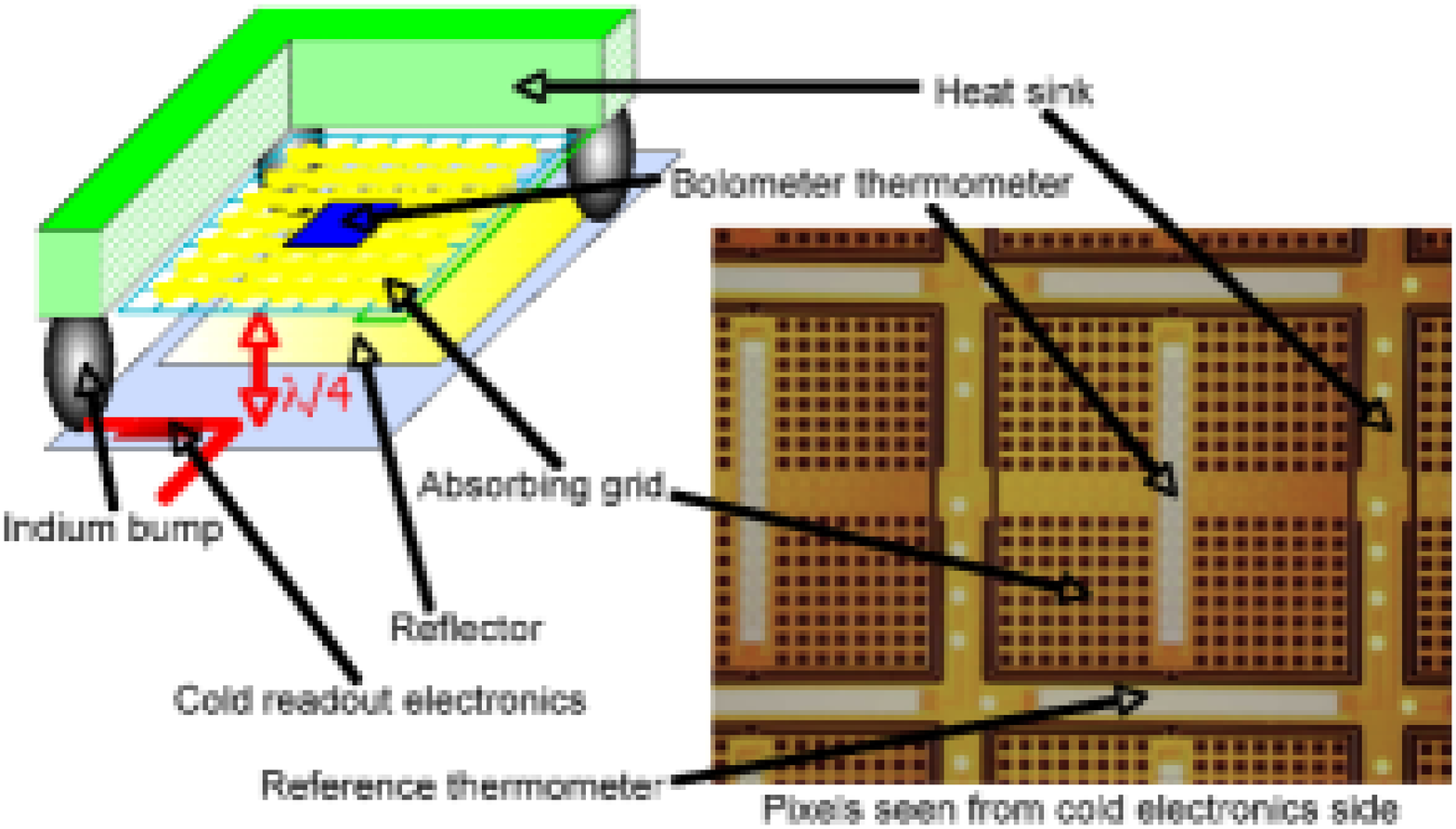}
\end{tabular}
\caption{\label{fig:structure} Structure of a bolometer pixel. Pixel
step is 750~$\mu$m.}
\end{center}
\end{minipage}
\hfill
\begin{minipage}{0.38\linewidth}
\begin{center}
\begin{tabular}{c}
\includegraphics[width=5cm]{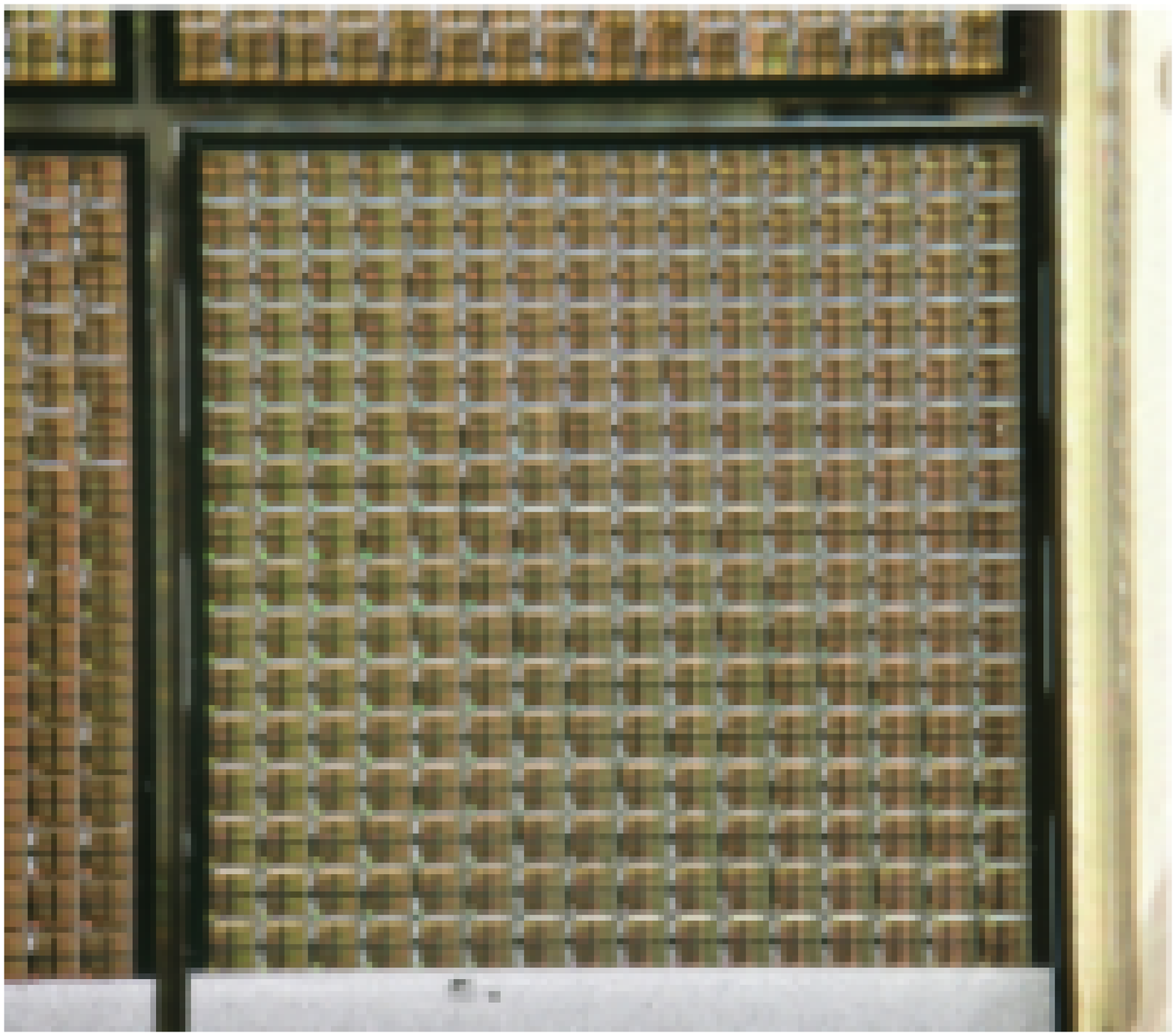}
\end{tabular}
\caption{\label{fig:array}View of an array integrated in the Blue
focal plane showing 256 multiplexed bolometers. Array dimensions are
12.63$\times$15.78~mm.}
\end{center}
\end{minipage}
\end{figure}

Figure~\ref{fig:structure} shows the structure of the bolometer pixel,
while figure~\ref{fig:array} shows a close-up view of a bolometer
array. More details on the technology of these bolometer arrays can be
found in reference~\citenum{simoens_1}.

\subsection{Irradiations} 
\label{sect:irradiations}
A ``total ionizing dose'' measurement was performed using a $^{60}$Co
gamma ray source. The purpose was to observe any damage due to protons
and electrons after an irradiation level equivalent to the expected
cumulated ionizing dose at the end of the mission (11 krads). No
significant degradation, either in thresholds or gains of the
bolometer arrays, has been observed, and the detectors can withstand
the spatial environment without degradation of their performances.

\begin{figure}[htbp]
\begin{center}
\begin{tabular}{c}
\includegraphics[height=7cm]{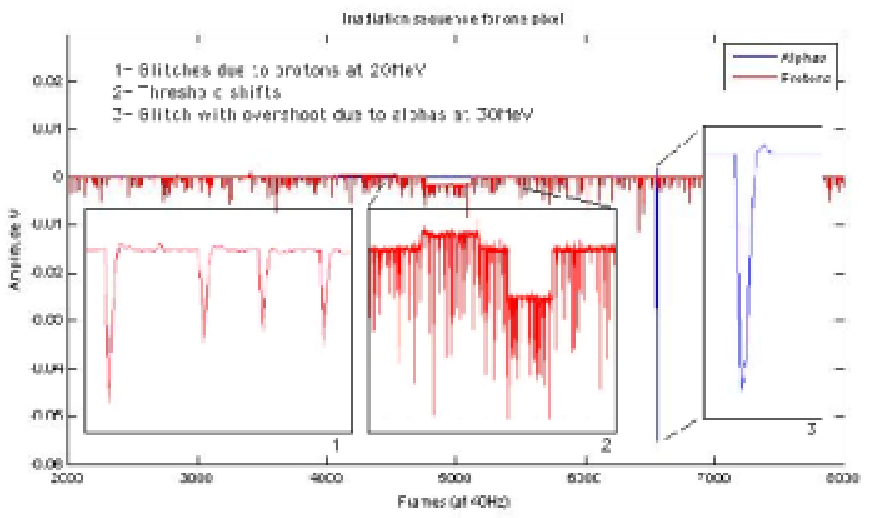}
\end{tabular}
\end{center}
\caption
%>>>> use \label inside caption to get Fig. number with \ref{}
{\label{fig:irradiations} Different effects of alpha and proton
impacts on the bolometers.}
\end{figure}

Another irradiation run was performed to look for single event
effects: proton and alpha irradiations were performed at the
Institut de Physique Nucl\'{e}aire (IPN, Orsay, France), respectively
at 20~MeV and 30~MeV with fluences of $\sim$3 particle/sec/pixel and
$\sim$0.2 particle/sec/pixel on a dedicated bolometer array
representative of the flight model. We observed no significant
variation of the gain of the detector. The main effects were threshold
shifts and glitches (see figure~\ref{fig:irradiations}). Threshold
shifts can be explained by the passage of particles through the CMOS
cold readout electronics located just below the detection
layer. However most of the perturbations were glitches due to the
passage of ions through the absorbing mesh resulting in a temperature
increase of the pixel. These caused a rapid signal variation, with an
average duration of about 4 frames (at a frame rate of 40~Hz) and an
amplitude ranging from 1~mV to 60~mV with a mean value of about
10~mV. A preliminary analysis shows a spatial distribution of the
incident particle signatures (number of frames affected, or relaxation
time) correlated with the responsivity map of the pixels (or the
bolometer impedances at first order).

%%-----------------------------------------------------------
\section{The cryocooler} 
\label{sect:cryocooler}
The SPIRE and PACS instruments on-board the Herschel spacecraft have
the same sorption coolers, which are based on an $^3$He evaporative
cooling cycle. The units have been designed following the same overall
specifications (see table~\ref{tab:coolers_1}). They feature the same
mechanical interface and if necessary can be swapped with a few minor
adjustments.

\begin{table}[htbp]
\caption{\label{tab:coolers_1}Herschel cooler specifications.}
\begin{center}       
\begin{tabular}{|l|l|}
\hline Safety & Structural failure mode shall be leak before burst \\
\hline & Sine sweep vibration: 22.5 G peak up to 100 Hz \\
Mechanical & Random 20 -- 150 Hz: 11.5 G rms \\
(worst case axis) & First eigenfrequency above 120 Hz \\
& Proof pressure: 2 x maximum operating pressure \\
\hline & Heat lift capability: 10 $\mu$W minimum at 290 mK \\
Thermal & 5 Joules of gross cooling energy at 300 mK \\
& Recycling time: no more than 2 hours \\
& Total energy dissipated per cycle: no more than 860 J \\
\hline Electrical & Cold interface (cooler heart) electrically insulated from mechanical interfaces \\
\hline Geometry & Volume and Mass: 100x100x230 mm maximum -- $ <$ 1.8 Kg \\
and Interface & Mechanical interface: with a 4 K structure \\
 & Thermal interface: with a 1.7 K $^2$He bath \\
\hline
\end{tabular}
\end{center}
\end{table} 

\begin{table}[htbp]
  \caption{\label{tab:coolers_2}Cooler main characteristics.} 
  \begin{center}       
    \begin{tabular}{|l|l|}
      \hline
      He charge & $\approx$ 6 STP dm$^{3}$ \\
      \hline
      Pressure at room temperature & $\approx$ 8.4 MPa \\
      \hline
      Overall dimensions & 100x100x229 mm \\
      \hline
      Overall mass & 1750 grams \\
      \hline
      Suspended mass (cooler "heart") & 280 grams \\
      \hline 
    \end{tabular}
  \end{center}
\end{table} 

The thermal architecture in the satellite is such that the coolers are
mechanically mounted off a structure at 4 K or above (``level 1'') and
thermal paths are then provided to the superfluid tank (``level 0'') for
the cooler operation. This constraint calls for a specific thermal
architecture and design. In addition during cooler operation, in
particular during the recycling phase, the heat flows to the tank from
the sorption pump and from the evaporator are significantly
different. During this phase it is crucial to keep the evaporator
temperature as cold as possible to increase the condensation
efficiency and reduce the fraction of liquid lost during
cooldown. Consequently two thermal interfaces and thus two thermal
buses to the superfluid tank are required. Finally to fulfil the
electrical insulation requirement, two gas gap heat switches are
mounted on the mechanical frame using Kapton spacers.

\begin{figure}[htbp]
  \begin{minipage}{0.34\linewidth}
    \begin{center}
      \begin{tabular}{c}
	\includegraphics[angle=270,width=3.5cm]{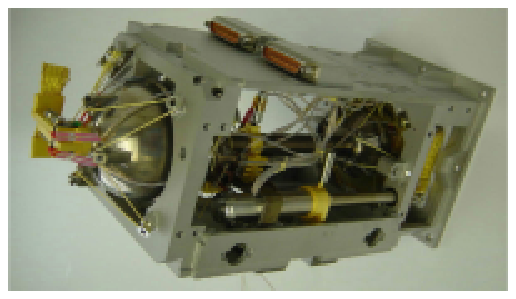}
      \end{tabular}
    \end{center}
    \caption{\label{fig:coolers}The PACS sorption cooler.}
  \end{minipage}
  \hfill
  \begin{minipage}{0.64\linewidth}
    \begin{center}
      \begin{tabular}{c}
	\includegraphics[height=7cm]{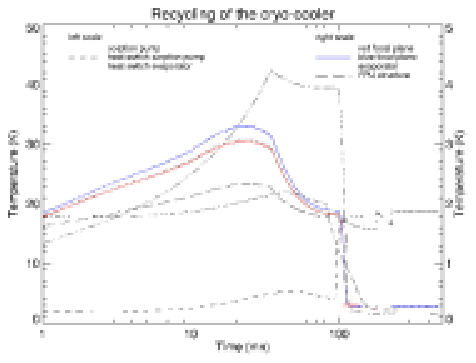}
      \end{tabular}
    \end{center}
    \caption{\label{fig:cycle}The first hours of a cooling cycle}
  \end{minipage}
\end{figure}

The hold time is one of the most critical performances for the SPIRE
and PACS instruments as any loss in autonomy can substantially impact
the mission and the amount of data expected. The autonomy of the
cooler in nominal operations is about 59
hours. Table~\ref{tab:coolers_2} gives the main characteristics of the
cryocooler, figures~\ref{fig:coolers} and \ref{fig:cycle} show a
picture of the cooler and a recycling performed in the integrated
flight model of the photometer. More details on the sorption cooler
can be found in ref.~\citenum{duband}. The lowest temperature achieved
at the evaporator for a bath at 1.6 K is 257.7 mK. The temperature of
the evaporator as measured in the PhFPU test cryostat is 283 mK for a
load of 5 $\mu$W (with all 10 arrays switched on).

%%-----------------------------------------------------------
\section{The warm electronics}
\label{sect:electronics}

\subsection{Overview of BOLC}
\label{sect:bolc_description}
The PACS Warm Electronics system comprises a unit called ``BOLC''
specifically in charge of the control of the photometer focal
planes. The main functions of this unit are (1) to act as the back end
for photometer instrumentation contained in the cryovessel and (2) to
interface with the rest of the PACS instrument. It includes
electronics associated with the bolometer arrays as well as
electronics associated with the thermal control of the bolometers
(temperature monitoring and control of the cryocooler).

The BOLC layout is based on electronics modules connected to a back
plane for digital communication between them. Electronics modules are
constituted of multi-layer circuit boards populated with SMD parts
assembled on individual chassis having the function of stiffener /
module holder / front panel connector fixation / thermal heat
sink. Additionally a separate specific enclosure contains the power
related function (Power Supply Unit).

BOLC dimensions are 382.5$\times$289$\times$333.5 mm and its weight is
18.25 kg. The power budget of the warm electronics is 44.2 W in
nominal operation mode (i.e. observing), 6.9 W during recycling and
6.0 W in stand-by.

\subsubsection{Analog Signal Processing}
\label{sect:bolc_analog}
\begin{figure}[htbp]
\begin{center}
\begin{tabular}{c}
\includegraphics[width=12cm]{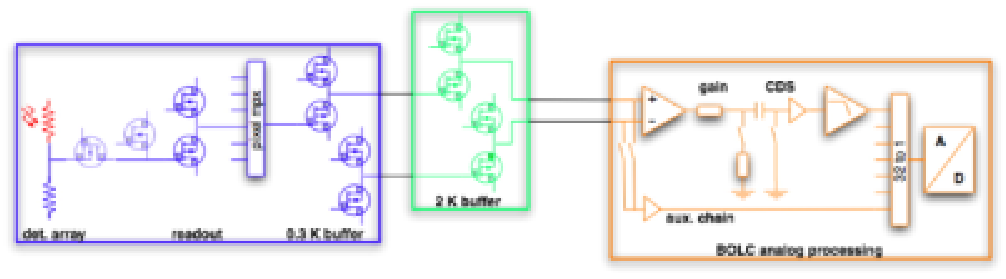}
\end{tabular}
\end{center}
\caption{\label{fig:cold_electronics}The PhFPU analog electronics.}
\end{figure}

The analog signal processing chain is divided into several stages: two
are part of the detector assembly: the cold readout electronics (at
300~mK) and the cold buffer (at 2~K), as shown in
figure~\ref{fig:cold_electronics}.

The last stage (at warm temperature) is within the BOLC unit. The
instantaneous dynamic range is determined by the analog to digital
converter limited by cold electronics transfer function mismatches: a
value of 65000 can be considered. The full dynamic range takes
advantage of the gain switching and of the detector noise
over-sampling and reaches a value of 330000 (corresponding to the
range from 3 mJy to 1000 Jy for incident optical flux). A total number
of 5 modules is required to process the 160 bolometer signal outputs:
4 dedicated to the Blue focal plane and 1 to the Red focal
plane. Limited power budget for the whole unit implies the design of
very low power analog channel while keeping low noise performance:
measurement on the flight model gives a value of 9.4 mW per readout
channel.

\subsubsection{Detector clock sequencer and bias generator}
\label{sect:bolc_sequencer}
BOLC contains adjustable voltage generators and clock translators
required to bias and address a bolometer array. All the parameters are
controlled by means of low level commands via digital module allowing
for optimization of the bolometer performance, according to initial
cold electronics parameter dispersion, detector illumination and
parameter drift during instrument life.

A total of 3 modules is required to handle independently 4 groups of 2
arrays for the Blue focal plane and 2 groups of 1 array for the Red
focal plane respectively, corresponding to the setting of more than
100 parameters, no less than 19 biases and clocks being required for
each detector. The module implementation includes 12-bit digital to
analog converters for adjustable settings, analog switches for on/off
functions while the digital functions are embedded into a single
radiation tolerant FPGA (RT1425 from ACTEL).

Clock translators are driven by a programmable sequencer: timing can
be trimmed by means of telecommands to optimize detector
performance. An additional signal is provided to the rest of the
instrument to achieve the synchronization of the PACS chopper with the
bolometer frame readout. The sequencer is implemented in the FPGA
along with other functions of the digital module. To perform quick
interface checking, the sequencer also features an internal pseudo
random data generation allowing data generation even if unplugged from
the PhFPU cold electronics.

\subsubsection{Ultra low temperature measurement}
\label{sect:bolc_temperature}
BOLC also controls the cryocooler and the acquisition of the
housekeeping parameters, including measurements of the photometer
temperature. Detector and cryocooler operations require various PhFPU
temperatures to be monitored and the measurements cover a range from
0.2 K to 50K.  Lower temperature measurements (0.2 K to 1 K) require a
very high resolution (0.0001 K).  For such measurements the probe bias
must be chosen to maximize sensitivity and to limit self heating
(Pprobe $<$ 1 nW). Even higher resolutions (10 $\mu$K) can be achieved
by accumulating samples over one second.

\subsubsection{SpaceWire digital interface}
\label{sect:bolc_digital}
The digital module handles analog module communications as well as
external communication with the PACS warm electronics based on a
single Command and Data interface running over the SpaceWire standard.
The SpaceWire core has been developed in order to optimize design
integration. Thus all the digital functions of the BOLC unit have been
embedded into a single radiation tolerant RT54SX32S FPGA from ACTEL .

\subsubsection{Redundancy}
\label{sect:bolc_redundancy}
Standard safety considerations have been taken into account in order
to deal with electronics failures: single point failures are avoided
and failure propagations are minimized. Unit internal redundancy relies
on both cold and warm redundancies: analog functions are shared into 6
independent modules each being devoted to a bolometer sub-assembly
(warm redundancy) while digital functions (clock sequencer, interfaces
to PACS instrument, internal interfaces to analog functions) are
doubled (cold redundancy). Therefore a failure at analog electronics
level is limited to a portion of the field of view and recovery from a
failure in digital electronics is simply achieved by switching from
main to redundant module.

%%-----------------------------------------------------------
\section{Performances of the Photometer Focal Plane Unit}
\label{sect:phfpu}

Delivery of the Photometer to the PACS consortium at the Max Planck
Institute for Extraterrestrial Physics (Germany) is scheduled for June
2006. The results presented in this section were obtained in CEA
Saclay during the first quarter of 2006 while preparing for the
calibration campaign and deal mainly with performance optimizations.

\subsection{Measuring RR ratios}
\label{sect:RR}

\begin{figure}
\begin{center}
\begin{tabular}{c}
\includegraphics[width=11cm]{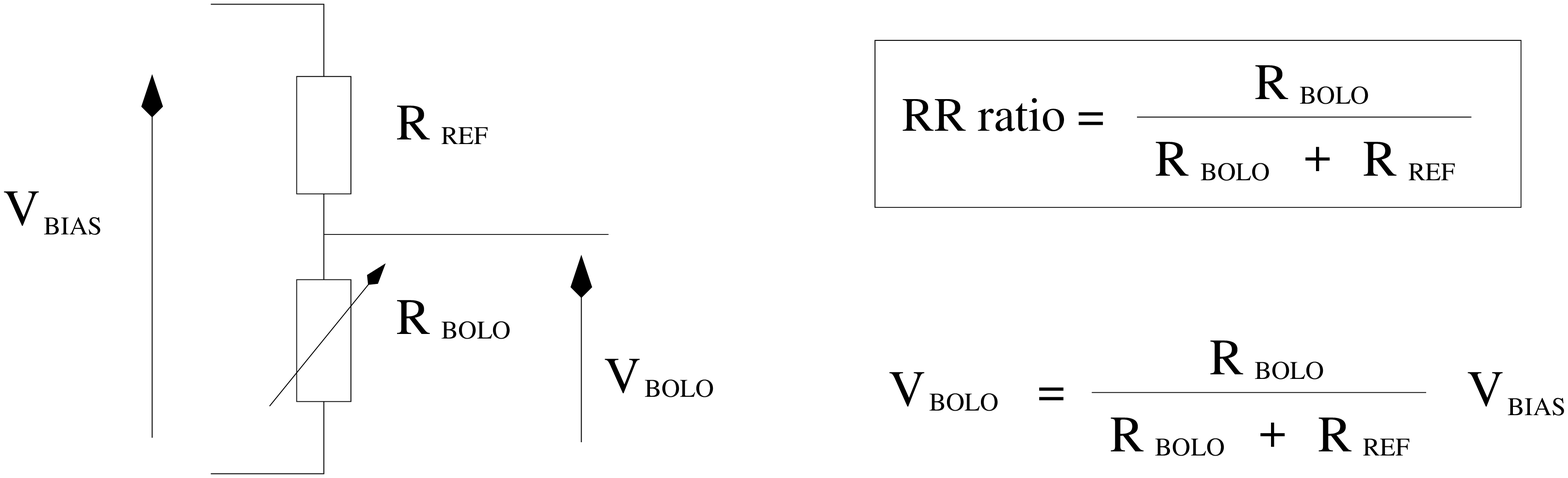}
\end{tabular}
\caption{Electrical setup of the bolometric bridge for a single pixel
(left). Definition of the RR ratio and the corresponding expression
for V$_{BOLO}$ (right).}
\label{fig:bolo_bridge}
\end{center}
\end{figure}

The bolometric signal, $V_{BOLO}$, is set at the middle point of a
voltage divider, also called bolometric bridge. It consists of a
bolometer resistor $R_{BOLO}$ and a reference resistor $R_{REF}$.
Figure~\ref{fig:bolo_bridge} shows the electrical setup of the
bolometric bridge and the definition of the \emph{RR ratio}. Both
resistors are strictly identical\footnote{Same doping, same size
(40$\times$600 $\mu m$) and impedance of about 0.7~T$\Omega$ at
300~mK}, however the bolometer's temperature, hence its impedance, is
allowed to fluctuate with the pixel temperature whereas the reference
resistor is in thermal contact with the inter pixel wall which acts as
the 300~mK heat sink (figure~\ref{fig:structure}). In this
configuration the reference resistor is used as a current source for
the bolometer and prevents any burnout phenomenon. Note that typical
voltages applied across the resistors are of the order of a few Volts
which induces electric field effects in the conduction process
(impedance decreases with applied voltage).

While most bolometers can be fully characterized by measuring their
load curves, we rather use \emph{RR ratios} to determine the state of
the bolometric bridge. Indeed the bolometers are multiplexed and each
array contains 256 bolometers mounted in parallel so we cannot measure
the current flowing through each bolometer individually. Moreover the
300~mK reference resistor impedance varies with the applied bias due
to electric field effects and it cannot be considered as an actual
load resistor.

RR ratios are obtained from static measurements with a stable and
thermalized black body illuminating the whole focal plane. We record
the 40~Hz-sampled signal of each of the 2560 bolometers for 3 minutes
in different bias/background flux configurations. In fact each of
these configurations requires a different tuning of the detectors to
avoid saturation of the signal. Reconstructing RR ratios by taking
into account the different offsets and gains of the electronics chain
provides us with absolute measurements of the bolometric bridge and
makes data obtained in very different conditions comparable. The
computation of RR ratios played an important role in understanding the
functioning of the bolometer arrays. RR ratios and their
corresponding $V_{BOLO}$ values are plotted in
figure~\ref{fig:rrratio}. RR ratios can be interpreted as a
competition between the reference and the bolometric resistors and
reflect the balance or rather the unbalance of the bolometric bridge.

\begin{figure}  
\begin{center}
\begin{tabular}{ll}
\includegraphics[width=8cm]{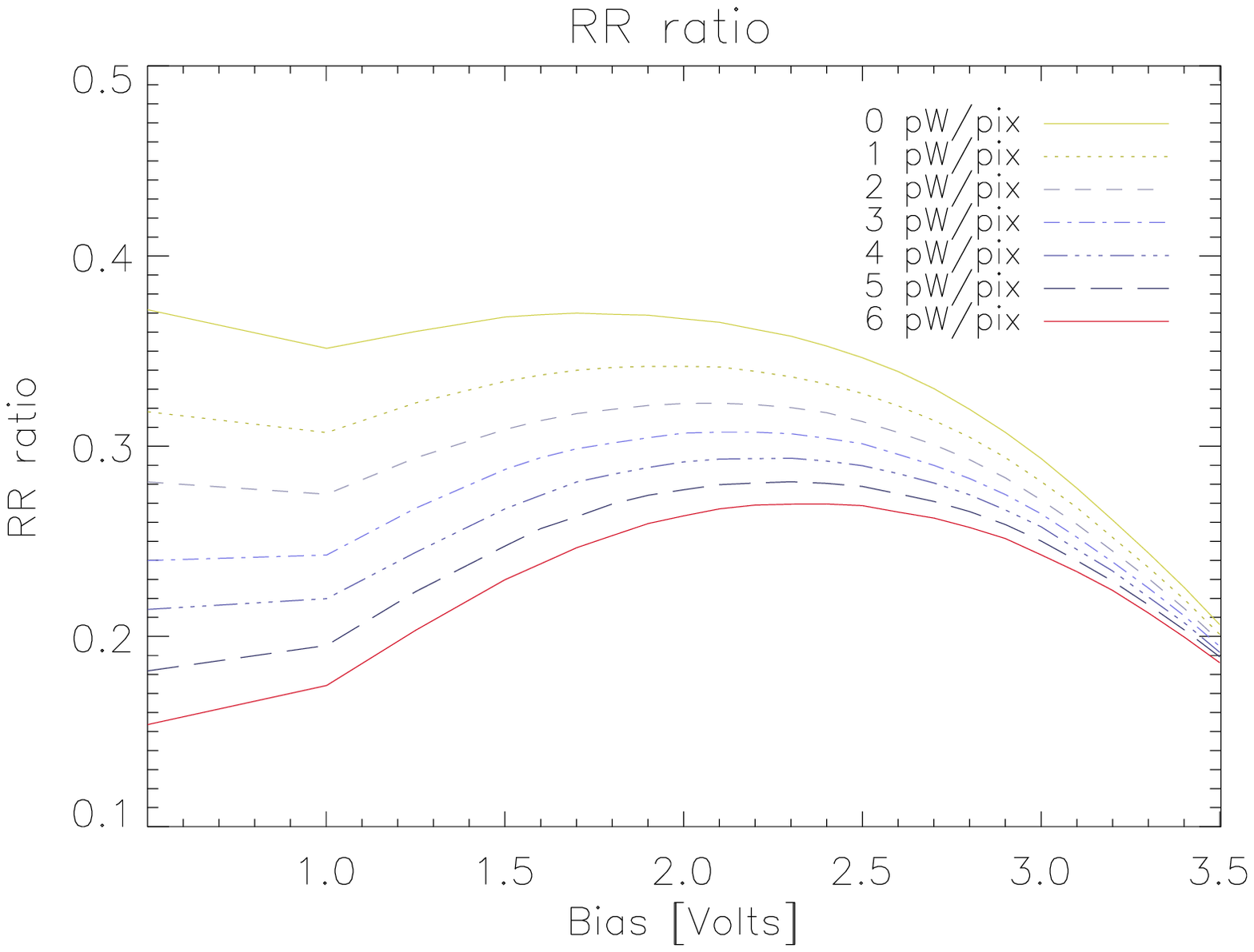}  &
\includegraphics[width=8cm]{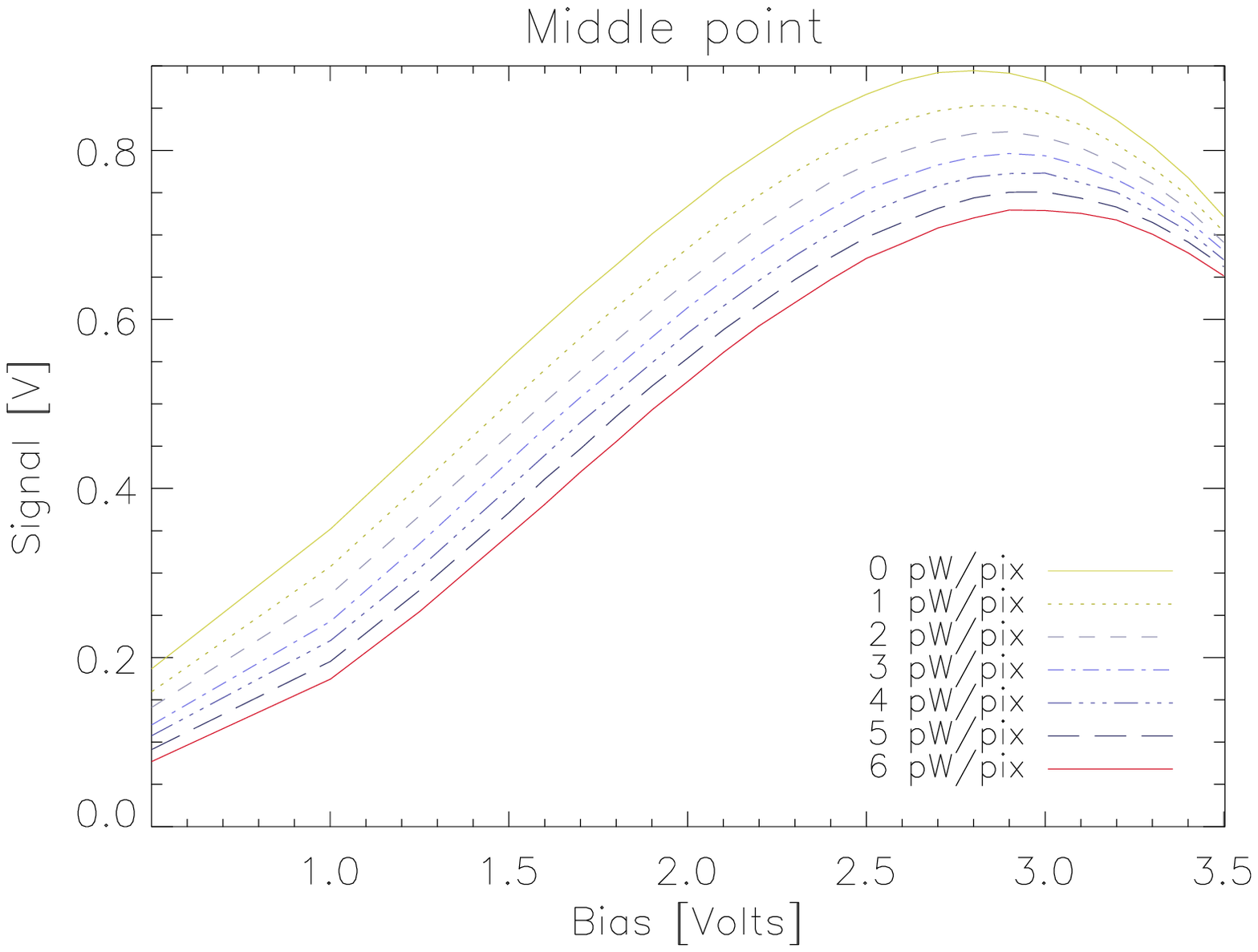}
\end{tabular}
\caption{Median \emph{RR ratios} of a single array plotted versus
voltage bias for different background fluxes (left) and the
corresponding $V_{BOLO}$ values (right). Top curves corresponds to
0~pW background and bottom curves to 6~pW.}
\label{fig:rrratio}
\end{center}
\end{figure}

RR ratio measurements provide us with a large and valuable amount of
information. For instance detector responsivity can be derived from
the data presented here as described in section
\ref{sect:resp}. Moreover we made an extensive use of these ratios to
predict the different voltages necessary to power up the
detectors\footnote{There is a set of 19 inter-dependant voltages
necessary to tune the bolometers of a single array}. Indeed the middle
points being quite dispersed, it is crucial to fine-tune the detectors
to ensure all pixels fit in the dynamics of the ADC.

\subsection{Responsivity measurements and non-linearity}
\label{sect:resp}

The responsivity is usually measured by modulating the incident flux
with a chopper, the responsivity is then the ratio of the signal
amplitude to the flux modulation amplitude. It is expressed in V/W. In
our case since RR ratios are reconstructed for different biases and
background fluxes we compute $\partial Signal/\partial Flux$ for each
pixel and for each flux and bias. Figure~\ref{fig:resp_map} shows a
responsivity map of the Blue focal plane for a bias of 2.2~V with a
background flux of 2~pW/pixel. The responsivity is about 4$\times
10^{10}$~V/W and is quite homogeneous over the whole focal plane.

Table~\ref{tab:resp_tab} presents the average responsivity of the Blue
focal plane for fluxes in the range 0 to 7~pW/pixel\footnote{We expect
a background flux from the telescope between 1 and 6~pW/pixel.} along
with its associated deviation. It shows only a slight non-linearity
over this background flux range.

\begin{figure}[htbp]
  \begin{minipage}{0.58\linewidth}
    \begin{center}
      \begin{tabular}{c}
	\includegraphics[width=9cm]{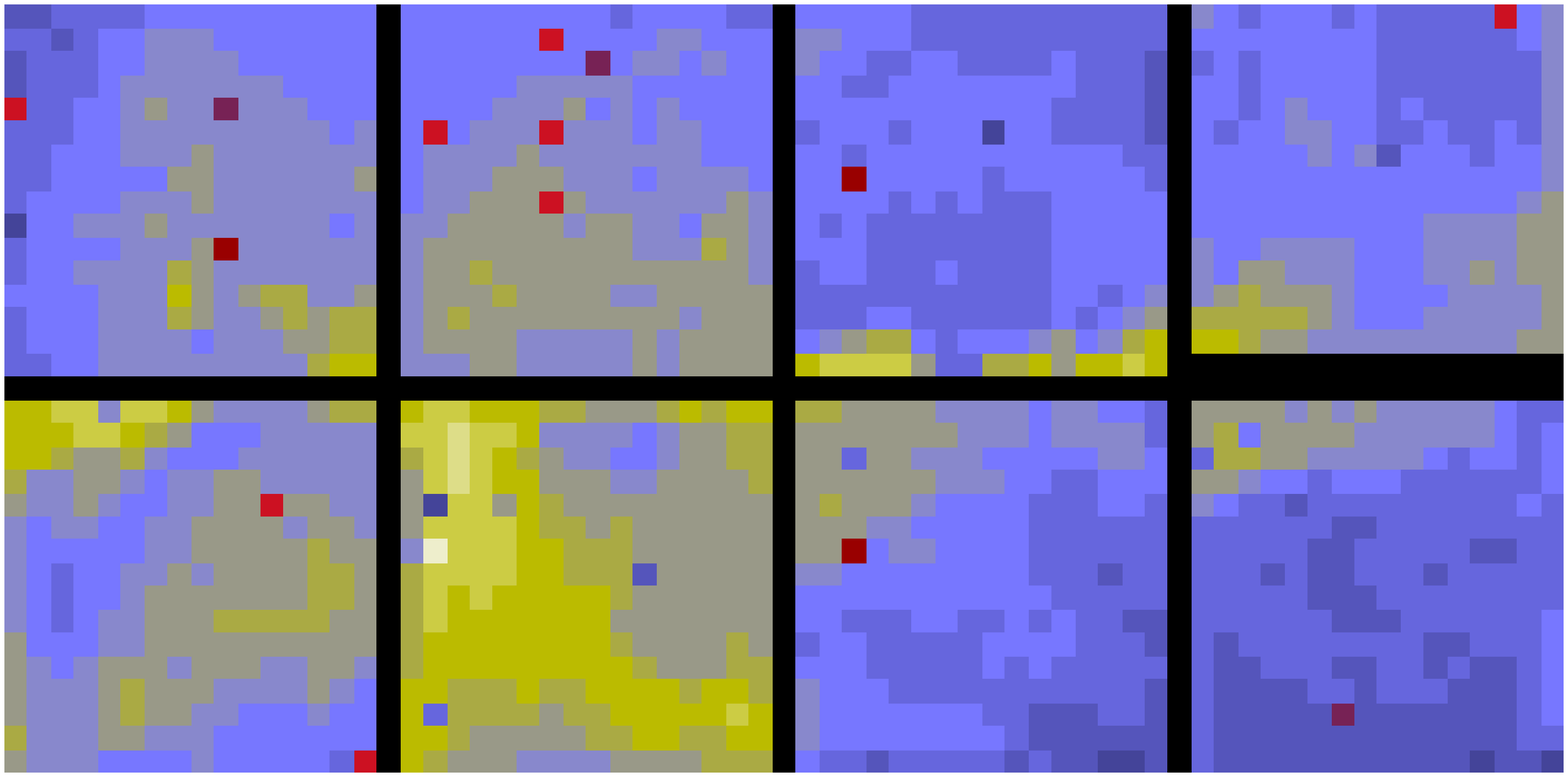}
      \end{tabular}
    \end{center}
    \caption{\label{fig:resp_map}Responsivity map of the Blue focal
      plane for a background flux of 2~pW/pixel and a bias of 2.2~V.}
  \end{minipage}
  \hfill
  \begin{minipage}{0.40\linewidth}
    \begin{center}
      \begin{tabular}{|c|c|c|}
	\hline
	Background  & Mean Resp.  & $\sigma$ \\
	Flux  (pW/pix) & ($\times$10$^{10}$ V/W) & (\%) \\
	\hline
	0 & 4.20 & 20.54 \\
	\hline
	1 & 4.18 & 20.49 \\
	\hline
	2 & 4.15 & 20.44 \\
	\hline
	3 & 4.13 & 20.39 \\
	\hline
	4 & 4.11 & 20.34 \\
	\hline
	5 & 4.08 & 20.29 \\
	\hline
	6 & 4.06 & 20.24 \\
	\hline
	7 & 4.03 & 20.18 \\
	\hline
      \end{tabular}
    \end{center}
    \renewcommand\figurename{Table}
    \addtocounter{figure}{-8} 
    \caption{\label{tab:resp_tab}Mean responsivities of the Blue focal
      plane measured in static mode for different fluxes and its
      associated relative dispersions.}  
    \addtocounter{table}{1}
    \addtocounter{figure}{+7} 
    \renewcommand\figurename{Figure}
  \end{minipage}
\end{figure} 

\subsection{Spectral noise density and bandpass cutoff frequency}
\label{sect:noise}

\begin{figure}[htbp]
\begin{center}
\begin{tabular}{cc}
\includegraphics[width=8cm]{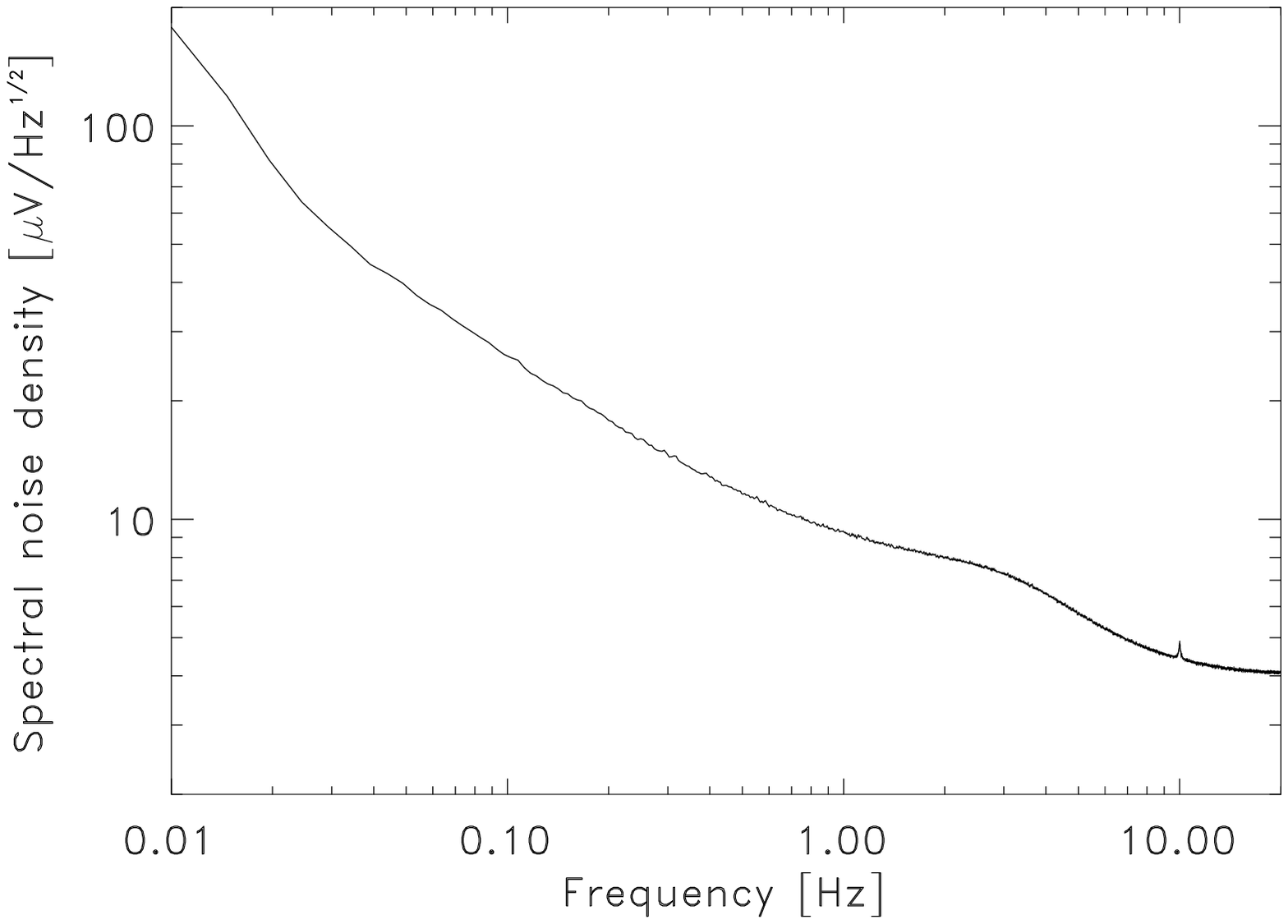} &
\includegraphics[width=8cm]{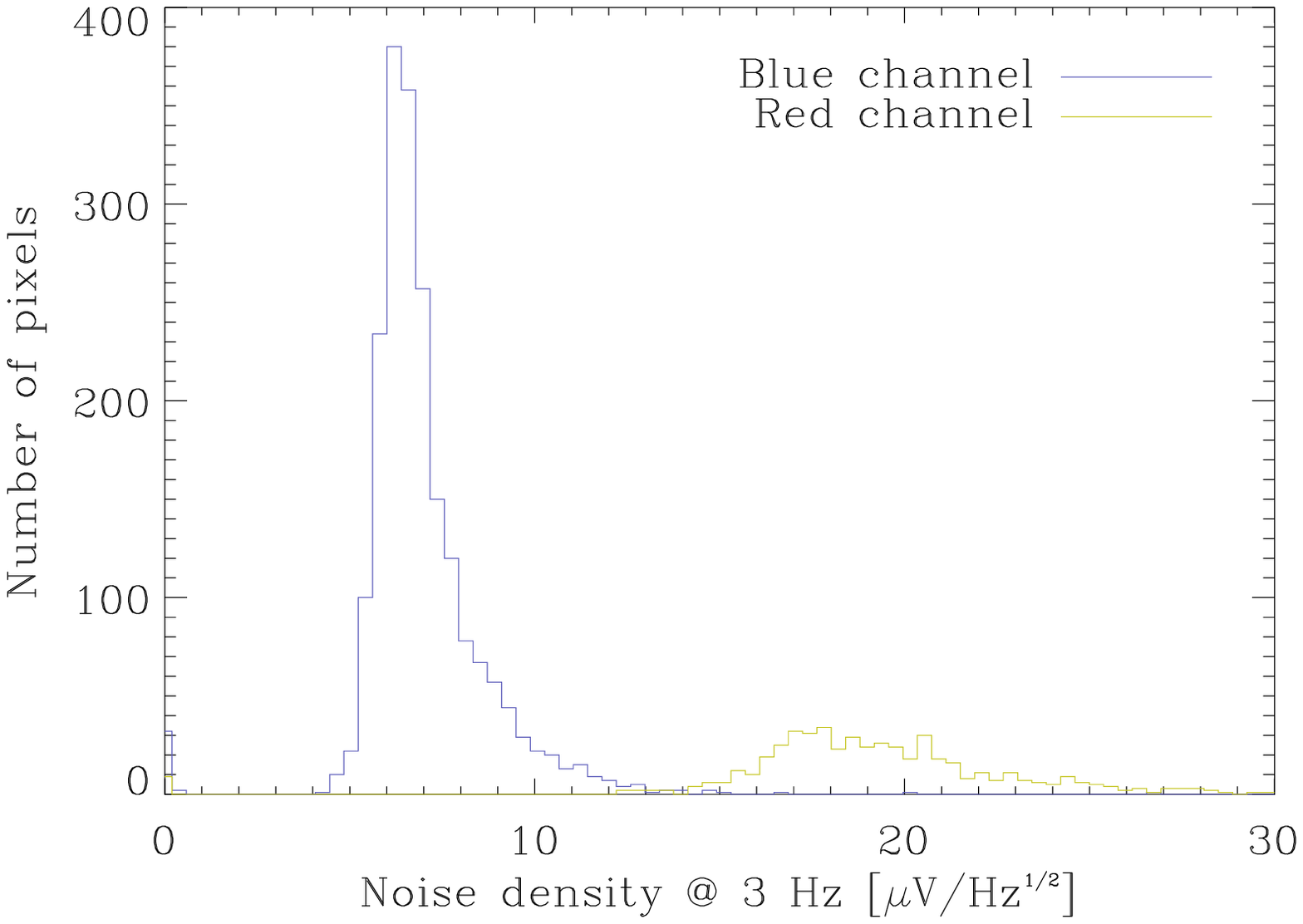}
\end{tabular}
\end{center}
\caption{\label{fig:oofnoise}Average noise spectrum of 256 pixels
from the same array (right). The histogram showing the noise
dispersion on the Blue and Red focal planes.}
\end{figure}

PACS bolometers exhibit noise levels two orders of magnitude higher
than ``usual'' bolometers. This is mainly due to the MOS cold readout
electronics located at 300~mK but also to their impedance close to
1~T$\Omega$. We actually measure noise levels at 3~Hz of about
7~$\mu$V.Hz$^{-1/2}$ on the Blue focal plane and 18~$\mu$V.Hz$^{-1/2}$
on the Red one. Figure~\ref{fig:oofnoise} shows the spectral noise
density extracted from a 3~hours measurement as well as a histogram of
noise levels measured at 3~Hz.  The spectrum was obtained by coadding
spectra of 4-minutes sub-samples to decrease the statistical
fluctuations. Moreover 256 spectra from pixels of the same array have
been averaged to obtain the final spectrum representative of the whole
array.

We distinguish 3 domains in this spectrum: (1) the low frequency
region ($f_{knee}<$0.5~Hz) concentrates most of the energy and is
responsible for signal drifts, (2) the operational regime between 0.5
and 5~Hz which is dominated by photon noise and (3) the white noise
filtered by the bolometers electrical time constant
($\tau\sim$60~ms). Additional bandpass tests confirmed the 5~Hz value
for the electrical bandpass cutoff frequency\footnote{The bandpass
cutoff frequency is defined as the modulation frequency at which the
signal is attenuated by 3~dB.} and revealed a thermometric time
constant of 20~ms. These time constants can possibly be lowered by
increasing the bias.

\subsection{Sensitivity measurements}
\label{sect:NEP}

In the sub-millimeter regime, the sensitivity is usually expressed as a
NEP (Noise Equivalent Power) and is defined as the ratio of the noise
level by the responsivity. We adopt this definition and we find an
optimal NEP of $2\times10^{-16}\,W/\sqrt{Hz}$ at 90 $\mu m$ and
$3.5\times10^{-16}\,W/\sqrt{Hz}$ at 160 $\mu m$ for a bias of
2~Volts at a background flux of 2~pW/pixel.

\begin{figure} [htbp] 
\begin{center}
\begin{tabular}{cc}
\includegraphics[width=8cm]{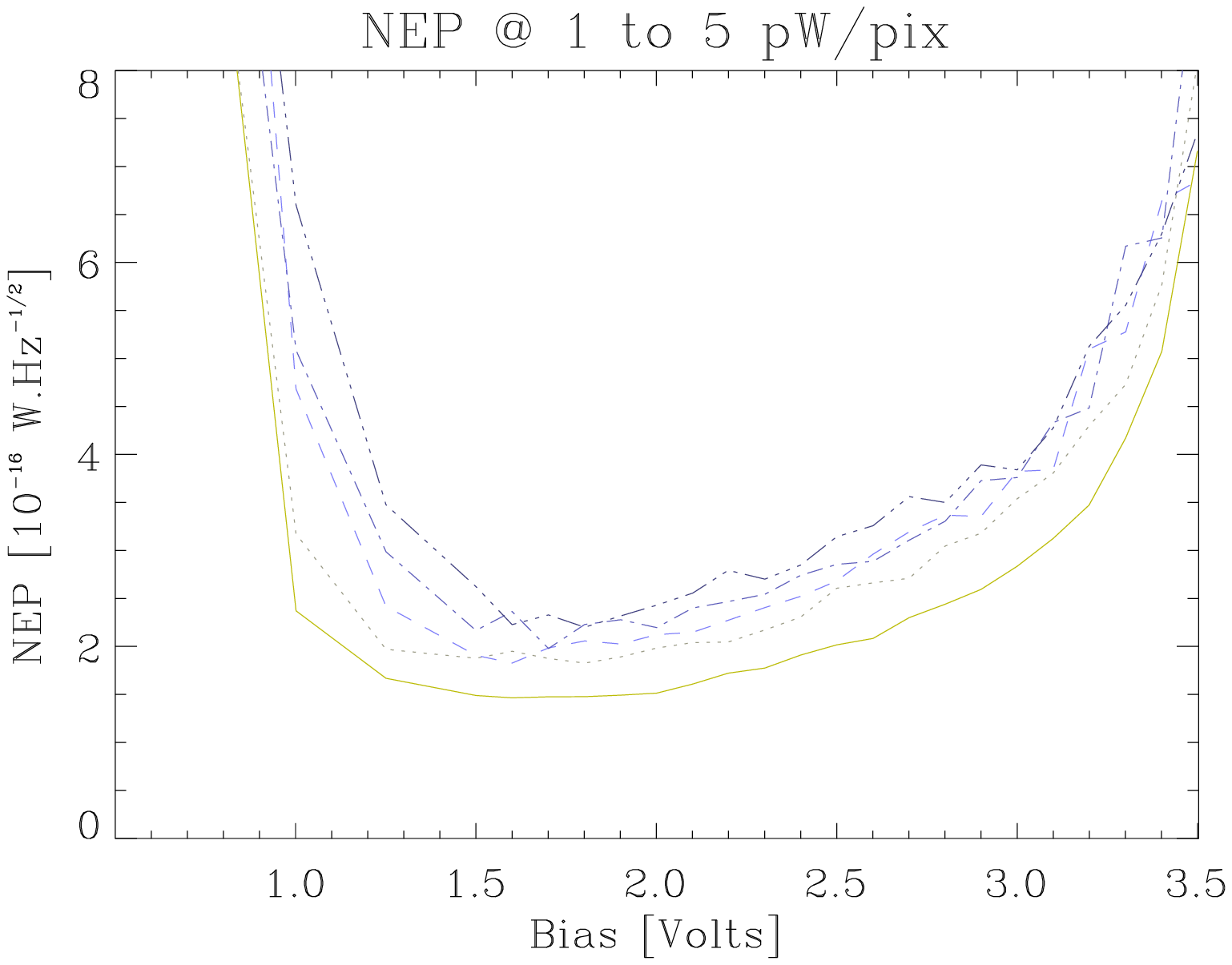} &
\includegraphics[width=8cm]{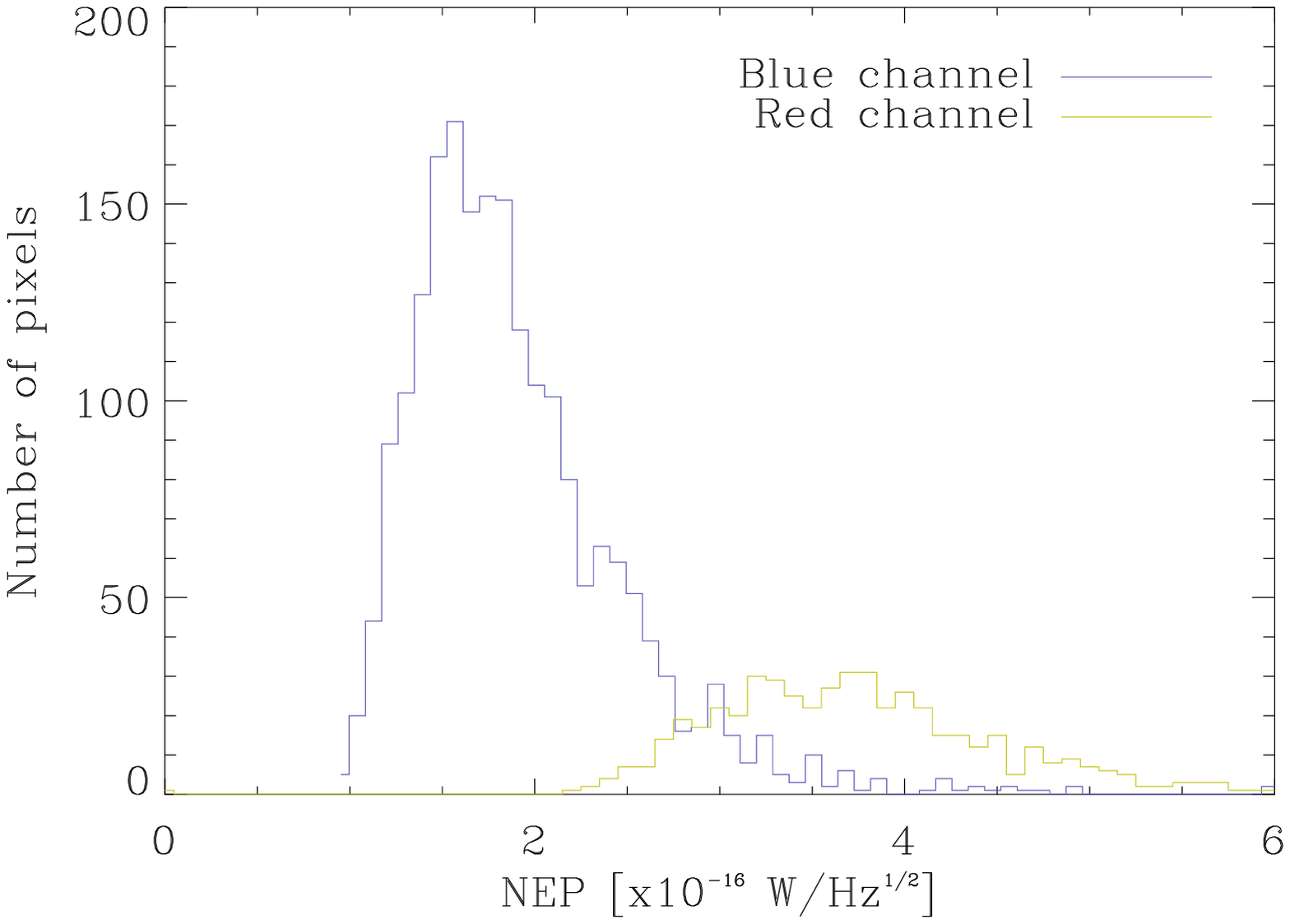} 
\end{tabular}
\caption{Measured median NEP obtained in static setup (without
chopping) for background fluxes of 1 to 5~pW/pixel from bottom to top
(left); the corresponding histogram showing the dispersion of NEP on
the whole PhFPU for a background of 2~pW/pixel and a bias of 2~V
(right).}
\label{fig:nep}
\end{center}
\end{figure}

Again, these results were derived from the data presented in
section~\ref{sect:RR}. The 3~minutes samples recorded in each
configuration were used to compute noise levels at 3~Hz and
responsivities were derived from $\partial Signal/\partial
Flux$. Figure~\ref{fig:nep} presents NEP values for fluxes between 1
and 5~pW/pixel as a function of the applied bias. The optimum NEP is
reached for a bias around 2~V independently of the background flux. At
low biases the Joule dissipation is small and the bolometers are too
cold, resulting in a very high impedance and thus high noise
levels. At higher biases, bolometers are heated up by Joule
dissipation which decreases significantly the impedance and thus the
responsivity.

%%-----------------------------------------------------------
\section{conclusion and future developments} 
\label{sect:futur}

The development of the bolometers for the Herschel/PACS instrument
demonstrates that it is now possible to build fairly large ($>$ 1000
pixels) focal planes based on filled arrays. The grid + resonant
cavity concept works, the science requirements on the Blue focal plane
have been reached and we will improve the performances of the Red
focal plane before delivery to the PACS consortium.

The future of this type of bolometers looks very bright. We have
already started a number of developments that will make use of these
detectors in the sub-mm and millimeter wavelength ranges, opening the
possibility of wide field imaging in these spectral domains. The two
main axes of development are (1) to redesign the packaging of the
arrays to make them 4-side buttable, opening the way for very large
focal planes of bolometers, and (2) to adapt the grid + cavity concept
for absorption at longer wavelength.

Table~\ref{tab:projets} lists the different projects in terms of
spectral range, size of focal planes, and telescopes.
\begin{table}[htbp]
\caption{\label{tab:projets}On-going developments of filled arrays of
bolometers.}
\begin{center}       
\begin{tabular}{|c|c|c|c|c|}
\hline Name & Spectral range & Focal plane & Telescope & Operational
in \\ \hline P-ARTEMIS & 2 channels 200 + 450 $\mu$m & 1 array per
channel & KOSMA + Chile & 2006 \\ PILOT & 2 channels 240 + 550 $\mu$m
& 2x2 arrays per channel & balloon & 2009 \\ ARTEMIS-1 & 3 bands 200
-- 350 -- 450 $\mu$m & 4x4 arrays & APEX & 2009 \\ ARTEMIS-2 & 3 bands
0.85 -- 1.2 -- 2 mm & 4x4 arrays & open (IRAM...) & 2010 \\ \hline
\end{tabular}
\end{center}
\end{table} 

The adaptation of the bolometers to longer wavelengths is simply a
tuning of the depth of the resonant cavity, in order to keep it at a
quarter of the wavelength of interest. This tuning can be done in
several ways: keeping the same cavity and putting a dielectric layer
with the proper thickness on top of the pixel, inserting a spacer in
between the absorbing grid and the reflector to adjust the depth of
the cavity, or a mix of these two methods. This adaptation to
longer wavelengths is presented in ref.~\citenum{reveret}, and the
ARTEMIS project is presented in greater details in
ref.~\citenum{talvard}.

\begin{figure}[htbp]
\begin{center}
\begin{tabular}{c}
\includegraphics[height=8cm]{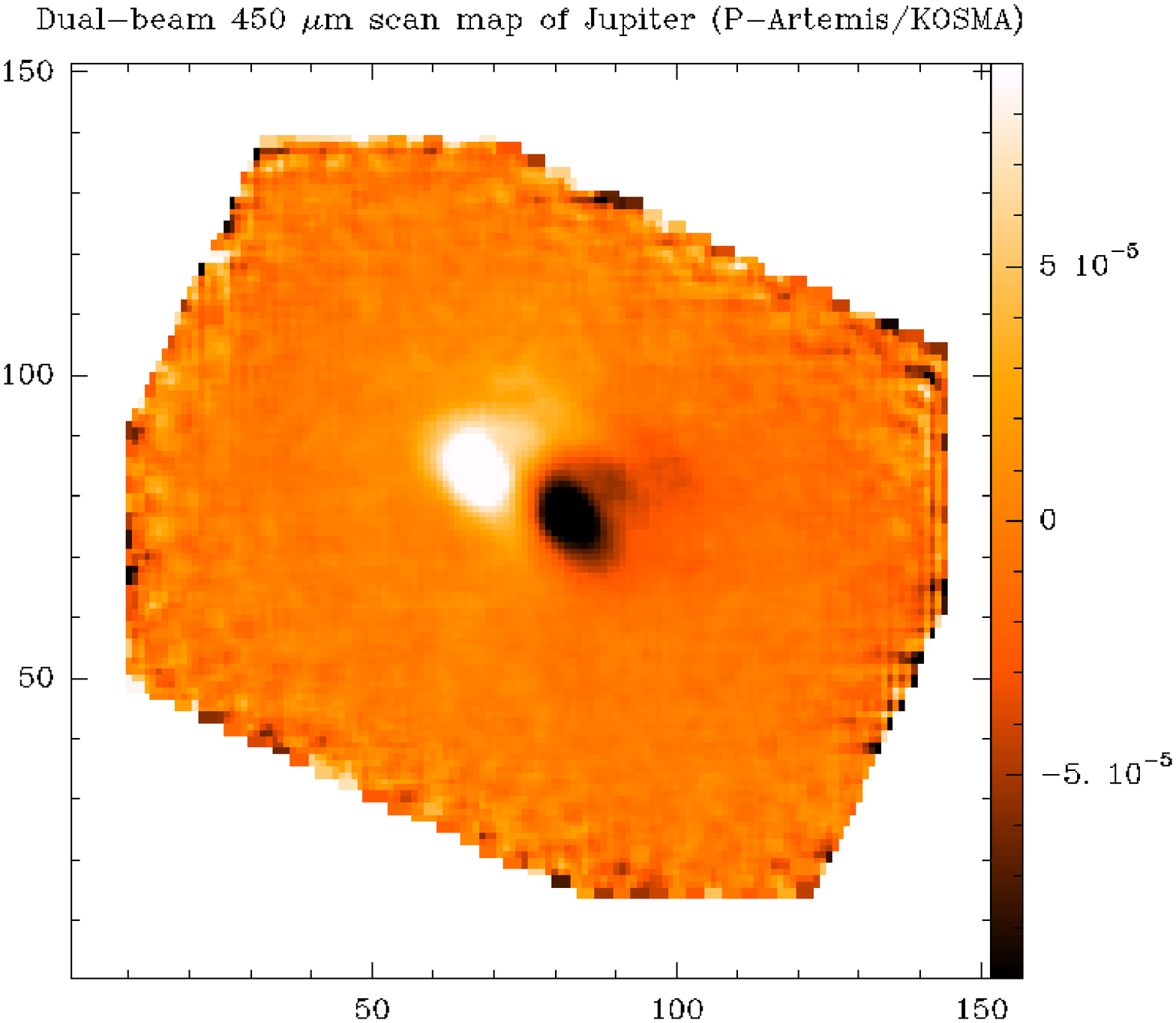}
\end{tabular}
\end{center}
\caption{\label{fig:kosma}Scanned map of Jupiter at 450 $\mu$m.}
\end{figure}

A first observing run of the P-ARTEMIS camera on the KOSMA telescope
occurred in March 2006, which demonstrated that these detectors
actually work at longer wavelengths. Figure \ref{fig:kosma} shows a
scanned map of Jupiter taken at 450 $\mu$m.

%%%%%%%%%%%%%%%%%%%%%%%%%%%%%%%%%%%%%%%%%%%%%%%%%%%%%%%%%%%%%

%\acknowledgments     %>>>> equivalent to \section*{ACKNOWLEDGMENTS}       

%%%%%%%%%%%%%%%%%%%%%%%%%%%%%%%%%%%%%%%%%%%%%%%%%%%%%%%%%%%%%
%%%%% References %%%%%

%\bibliography{report}   %>>>> bibliography data in report.bib

\bibliographystyle{spiebib}   %>>>> makes bibtex use spiebib.bst

\end{document}